\documentclass[prd,reprint,onecolumn,amsfonts,nofootinbib]{revtex4-2}
\usepackage{amsmath,amssymb}

\usepackage{slashed}
\usepackage{graphicx}
\usepackage{adjustbox}

\newcommand{\ii}{\mathrm{i}}
\newcommand{\dd}{\mathrm{d}}
\newcommand{\sgn}{\mathrm{sgn}\,}

\newcommand{\tg}{\tilde{\gamma}}
\newcommand{\tp}{\tilde{p}}
\newcommand{\tk}{\tilde{k}}
\newcommand{\kB}{\mathbf{k}}
\newcommand{\pB}{\mathbf{p}}
\newcommand{\qB}{\mathbf{q}}

\newcommand{\tr}{\mathrm{tr}\, }

\newcommand{\gam}{\mathfrak{g}}
\newcommand{\Dr}{\mathrm{Dr}}

\newcommand{\tm}{\mathrm{tm}}
\newcommand{\te}{\mathrm{te}}
\newcommand{\gr}{\mathrm{gr}}
\newcommand{\vf}{{v_F}}
\newcommand{\adb}{\allowdisplaybreaks } 
\newcommand{\ann}{\adb \nonumber \\}

\usepackage{xcolor} \definecolor{darkgreen}{rgb}{0,.5,0}
\usepackage[colorlinks,filecolor=blue,citecolor=darkgreen,unicode]{hyperref}
\begin{document}
\title{Impurities in graphene and their influence on the Casimir interaction}
\author{N. Khusnutdinov}\email{nail.khusnutdinov@gmail.com}
\author{D. Vassilevich}\email{dvassil@gmail.com}
\affiliation{Centro de Matemática, Computação e Cognição - Universidade Federal do ABC, Santo André, SP, Brazil}

\begin{abstract}
We study the influence of impurities in graphene described by a scattering rate $\Gamma$ on the Casimir interaction between graphene and an ideal conductor or between two identical sheets of graphene at zero temperature and chemical potential. To this end, we compute the polarization tensor of quasiparticles in graphene and corresponding conductivities for TE and TM channels. The Casimir energy density is evaluated with the help of the Lifshitz formula. We find that depending on the value of mass gap parameter the presence of $\Gamma$ may lead to a slight damping or to a considerable enhancement of the Casimir interaction.
\end{abstract}
\maketitle

\section{Introduction}\label{sec:intro}

The Casimir effect \cite{Bordag:2009zzd,Milton:2004ya} is an interaction of neutral bodies due to quantum vacuum fluctuations. The study of Casimir interaction between advanced materials is a new and promising area of research \cite{Woods:2015pla}. On one hand, the unusual electronic properties of these materials lead to interesting effects on the Casimir force. On the other hand, the improved quality of Casimir experiments makes them a useful tool for exploration of the materials themselves.

The Dirac materials (where the quasiparticles obey a quasirelativistic Dirac-type equation at sufficiently low energies) provide us with an example of interplay between Quantum Field Theory and Condensed Matter. Graphene is a prominent representative of this family \cite{CastroNeto:2007fxn,DasSarma2011}. Dealing with the Dirac materials it is natural to describe the interaction with electromagnetic field by the polarization tensor of quasiparticles and use this tensor to calculate the Casimir interaction. In the case of graphene, such an approach was used in \cite{Bordag:2009fz} and in \cite{Fialkovsky:2011pu} at zero and non-zero temperature, respectively. Remarkably, the polarization tensor approach to Casimir interaction of graphene was the only one which was confirmed at experiments \cite{Banishev:2013,Klimchitskaya:2014axa,Liu:2021ice,Liu:2021efo}.

All real materials contain impurities. Particular form of impurities can vary. Impurities refer to a general form for breaking the cleanliness of pristine materials. A classification of impurities and defects in graphene-like materials may be found in the reviews \cite{Banhart:2010:sdg,Zhang:2010:ig,Araujo:2012:diglm,Fritz:2013:pkig}. The two-dimensional nature of graphene decreases the number of possible types of defects and impurities. The point is that it is energetically favourable for adatoms or substitutional impurities to reside outside of the graphene surface. They may be charged \cite{Ando:2006:seismg,Cheianov:2006:foistdrg,Chen:2008:cisg}, magnetic \cite{Fritz:2013:pkig}, isotopic \cite{Fan:2003:mgcncil,Simon:2005:iecns}, topological such as pentagons and heptagons \cite{An:2001:sphclrstm,Zhang:2010:ig}, or be a consequence of imperfections and growth-induced defects like point \cite{Kotakoski:2006:eslrivtdcnas} and cluster defects \cite{Banhart:2010:sdg}. Intentional impurity is usually called a dopant while the impurity itself can be both intentional and unintentional (accidental). Doping is used to change the physical or chemical properties of a material. Impurities in graphene \cite{Altanhan:2012:ieg,Viola:2018:gpine} may transform a linear dispersion near Dirac points to a quadratic one which signifies the appearance of a mass gap induced by impurities. There are different approaches to describing impurities and their impact on the physical properties of materials. The common is a tight-binding model with a shot- or long-range potential \cite{Zhang:2010:ig}, and scattering approach \cite{Ando:1998:iscnabs,Ando:1998:bpabscn}.

With this large variety of the types of impurities in graphene we need a good model which captures the universal properties of impurities while being sufficiently simple for being used in calculation of the polarization tensor.  A successful way of describing impurities consists in adding to the propagator of quasiparticles a parameter $\Gamma$ which describes the impurity scattering rate. In other words, $\Gamma$ is an imaginary part of the fermion self-energy. Such a description has been applied to
graphene mostly in the presence of an external magnetic field in \cite{Gorbar:2002iw,Gusynin:2005iv,Gusynin:2006ym,Gusynin:2006gn,Fialkovsky:2012ee}. The computations of \cite{Fialkovsky:2012ee} are in a very good agreement with the measurements \cite{Crassee_2010} of giant Faraday rotation in graphene. In principle $\Gamma$ can depend on the frequency though keeping it constant appeared to be a good approximation. In this work, we neglect the other role of impurities which is their ability to create a non-zero chemical potential $\mu$.  

A particular form of impurities which were atoms (mostly sodium) on the surface of graphene and their influence on the Casimir force were considered in \cite{Liu:2021efo,Liu:2021ice}. According to these papers such impurities lead to a mass gap and a non-zero chemical potential of graphene but not to the appearance of impurity scattering described by scattering rate $\Gamma$. 

The primary goal of this paper is to study the influence of impurity scattering rate $\Gamma$ on the Casimir interaction between graphene and an ideal metal and between two graphene sheets. We restrict ourselves to the case of a zero temperature and vanishing chemical potential. This is a somewhat simplified setting. Therefore, we will not try to compare our results to any existing experiment (they were done at room temperature). Rather, we will clarify some generic features of the Casimir interaction of graphene. Our main conclusions is that turning on $\Gamma$ leads to a slight decrease of the Casimir energy density for small masses and to a considerable  enhancement thereof for large masses. Thus, taking impurities into account may be essential for analysing precision Casimir measurements.

This paper is organized as follows. In the next Section, we calculate the polarization tensor of quasiparticles in graphene in the Pauli--Villars subtraction scheme. In Section \ref{sec:Dru}, we analyse the conductivities following from this polarization tensor and discuss their properties which are relevant for the Casimir interaction. The Casimir interaction itself is studied in Section \ref{sec:Cas}. Section \ref{sec:concl} contains some concluding remarks. We put long formulas for the conductivities  in the Appendix. We use the units $\hbar=c=1$.
\section{Polarization tensor}\label{sec:pol}
To fix our conventions, we start with the Dirac operator
\begin{equation}
\slashed{D}=\ii \tg^i (\partial_i +\ii e A_i) - m, \label{DirOp}
\end{equation}
which describes the propagation of quasiparticles in graphene. The Latin letters form the middle of Alphabet will are used to denote coordinates of $2+1$ dimensional vectors, $i,j,k=0,1,2$. The letters from the beginning of Alphabet will denote spatial components, $a,b,c=1,2$. A twiddle above a vector means that the spatial components are rescaled with the Fermi velocity,
\begin{equation}
\tg^0=\gamma^0,\qquad \tg^a=v_F\gamma^a.\label{twiddle}
\end{equation}
The Dirac $\gamma$ matrices are $8$-dimensional (which corresponds to four generations of fermions in graphene). They satisfy the condition $\gamma^i\gamma^j+\gamma^j\gamma^i=2g^{ij}$ with the flat metric $g^{ij}=(+,-,-)$. Besides, $\tr (\gamma^i\gamma^j) = 8g^{ij}$ and $\tr (\gamma^i\gamma^j\gamma^l\gamma^k) =8(g^{i j}g^{lk} - g^{il}g^{jk} + g^{ik}g^{jl})$. We will use boldface to denote spatial components of the momenta 3-vectors, $k=(k_0,\kB )$, so that $p\cdot k=p_0k_0 - \pB\cdot \kB$ with $\pB\cdot \kB=k_a p_a\delta^{ab}$ with $\delta^{ab}$ being the Kronecker symbol.

The propagator $S(x)$ is the kernel of an inverse of a free Dirac operator $\slashed{D}_0=\slashed{D}_{A=0}$, i.e. $\slashed{D}^x_0 S(x-y)=\delta^{(3)}(x-y)$. After a Fourier transform 
\begin{equation}
	S (x) = \int \frac{\dd^3 k}{(2\pi)^3} e^{\ii kx} S(k),
\end{equation}
one arrives at the following expression
\begin{equation}
S (k) = - \frac{k_0 \gamma^0 + v_F k_a \gamma^a +m}{k_0^2 - v_F^2 k^2 -m^2}.\label{prop}
\end{equation}

To introduce impurities, characterized by the scattering rate $\Gamma$, and  chemical potential $\mu$ one shifts the temporal component of the momentum $k$ in the propagator (\ref{prop}) as $k_0\to \widehat{k}_0=k_0 + \ii \Gamma \sgn (k_0) +\mu$, (see \cite{Gorbar:2002iw,Gusynin:2005iv,Gusynin:2006gn,Gusynin:2006ym,Tkachov2013}) For the rest of this work we set $\mu=0$. We denote $\widehat{k}=(\widehat{k}_0,\kB )$.

The introduction of $\Gamma$ into the Quantum Field Theory approach leads to certain difficulties with the gauge invariance. The modification of fermionic propagator can be understood as a consequence of an additional term in the Dirac action containing $\psi^\dag \Gamma \sgn (-\ii \partial_0)\psi$. To maintain gauge invariance, each partial derivative has to be accompanied by a gauge field. That is, one should consider the expression $\psi^\dag \Gamma \sgn (-\ii (\partial_0+\ii e A_0))\psi$. Although it is not clear how one should deal with the sign function of a differential operator, new (and rather complicated) vertexes involving $\psi^\dag$, $\psi$, and $A_0$ seem to be inevitable. To overcome this difficulty we proceed as in \cite{Holanda:2023joy}. Namely, we consider only the diagrams without $A_0$ lines which are thus independent of whichever vertices involving $A_0$. I.e., we consider only the spatial components of polarization tensor,
\begin{equation}
	\Pi^{ab} (p) = \ii e^2 \int \frac{\dd^3k}{(2\pi)^3} \tr \left(S(\widehat{k}) \tg^a S(\widehat{k - p}) \tg^b\right). \label{Piab}
\end{equation}
The full tensor may be recovered, if needed, with the help of the transversality condition.

In the expression (\ref{Piab}), one can change the integration variable $k\to \tk$. The Jacobian factor $v_F^{-2}$ cancels $v_F^2$ coming from $\tg^a$ and $\tg^b$. As a result, the whole dependence on (\ref{Piab}) $v_F$ remains in the external momentum only, so that
\begin{equation}
\Pi^{ab}(p)=\Pi^{ab}_{v_F=1} (\tp).\label{vF1}
\end{equation}
In other words, to compute the spatial components of polarization tensor it is sufficient to do the computations for $v_F=1$ and then replace $p$ by $\tp$. Till the end of this section, all computations will be done for $v_F=1$.  Thus, 
\begin{equation}
	\Pi^{ab} (p) = \frac{\ii e^2 }{\pi^3}\int \dd^3k \frac{-\delta^{ab} (m^2 - \widehat{k}\cdot (\widehat{k - p})) +2 k^a k^b -  k^{a} p^{b}-k^bp^a}{(\widehat{k}^2 -m^2)((\widehat{k-p})^2 -m^2)}.
\end{equation}
By using the Feynman parametrization this integral can be rewritten as
\begin{equation}
\Pi^{ab} (p) = \frac{\ii e^2 }{\pi^3}\int \dd^3k \int_0^1\dd x \,\frac{-\delta^{ab} (m^2 - \widehat{k}\cdot (\widehat{k - p})) +2 k^a k^b -  k^{a} p^{b}-k^bp^a}{(x [\widehat{k}^2 -m^2]  + (1-x)[(\widehat{k - p})^2 -m^2])^2}.\label{Pab3}
\end{equation}
The denominator of the integrand reads
\begin{eqnarray}
&&(x [\widehat{k}^2 -m^2]  + (1-x)[(\widehat{k - p})^2 -m^2])^2=(\qB^2 -R)^2,\nonumber \\
&&\qB=\kB - (1-x)\pB,\\
&&R=x\widehat{k}_0^2 +(1-x)(\widehat{k-p})_0^2 -x(1-x) \pB^2-m^2.\nonumber
\end{eqnarray}
We change the integration variable $\dd^2\kB\to\dd^2\qB$. Note, that the terms in numerator of (\ref{Pab3}) which are linear in $\qB$ are integrated to zero. Also, under the integral one may replace $\qB^a\qB^b$ by $\tfrac 12 \delta^{ab}\qB^2$. After the integration over $\qB$ one arrives at the expression
\begin{equation}
\Pi^{ab} (p) = \frac{\ii e^2 }{\pi^2}\int_0^1\dd x \int \dd k_0 
\frac{\delta^{ab} (m^2 - \widehat{k}_0 (\widehat{k - p})_0-x(1-x) \pB^2) +2x(1-x) p^a p^b}{R}.
\end{equation}

To integrate over $k_0$ one has to split the integration region into a union of intervals where the signs of $k_0$ and $k_0-p_0$ are constant. If, for simplicity, we restrict ourselves to the case $p_0>0$ these are the intervals $(-\infty,0)$, $(0,p_0)$, and $(p_0,\infty)$. One can easily show that the integrals over the first and the last intervals coincide. We have,
\begin{eqnarray}
&& \Pi^{ab} (p) = \frac{\ii e^2 }{\pi^2}\int_0^1\dd x \left[ 2 \int_{-\infty}^0 \dd k_0\, \frac{\delta^{ab} (m^2 - ({k}_0-\ii \Gamma) ({k_0 - p_0}-\ii \Gamma)-x(1-x) \pB^2) +2x(1-x) p^a p^b}{x({k}_0-\ii \Gamma)^2 +(1-x)(k_0-p_0-\ii \Gamma)^2 -x(1-x) \pB^2-m^2}\right.\nonumber\\
&&\qquad\qquad\qquad +\left.\int_{0}^{p_0} \dd k_0\, \frac{\delta^{ab} (m^2 - ({k}_0+\ii \Gamma) ({k_0 - p_0}-\ii \Gamma)-x(1-x) \pB^2) +2x(1-x) p^a p^b}{x({k}_0+\ii \Gamma)^2 +(1-x)(k_0-p_0-\ii \Gamma)^2 -x(1-x) \pB^2-m^2}\right].\label{Piab4}
\end{eqnarray}

The integral over $k_0$ is divergent, so that we need a regularization and renormalization procedure for which we choose the Pauli--Villars subtraction. We subtract from the integrand in (\ref{Piab4}) the same expression due to a regulator field with a mass $M$, compute the integrals and then send $M\to \infty$. Symbolically, 
\begin{equation}
\Pi^{ab}_{\mathrm{PV}}(p)=\lim_{M\to\infty} (\Pi^{ab}(p,m)-\Pi^{ab}(p,M)).\label{PV}
\end{equation}

We represent the polarization tensor through two form factors, $\alpha$ and $\beta$, as
\begin{eqnarray}
\Pi^{ab}_{\mathrm{PV}}(p)=\alpha(p)\delta^{ab}+\beta(p)\, \frac{p^ap^b}{\pB^2}.\label{Piab5}
\end{eqnarray}
These form factors read
\begin{eqnarray}
&&\alpha(p)=\frac{2\ii e^2 }{\pi^2}\int_0^1 \dd x \left\{\frac{ -\ii \pi p^2 (1-x) x}{\sqrt{-q}} - \frac{2 p^2 (1-x) x }{\sqrt{q}} \arctan\left(\frac{p_0 x + \ii \Gamma }{\sqrt{q}}\right) + \frac{2 P^2 (1-x) x }{\sqrt{Q}}  \arctan\left(\frac{P_0 x - \ii \Gamma }{\sqrt{Q}}\right)\right.
\nonumber\\
&&\qquad\qquad +\left. \frac{1}{2} p_0 (1-2 x) \ln \left((p_0 x + \ii \Gamma )^2 + q\right) -  \frac{1}{2} P_0 (1-2 x) \ln \left((P_0 x - \ii \Gamma )^2 + Q\right)\right\},\label{albeta}\\
&&\beta(p)= -\frac{4\ii e^2 \pB^2}{\pi^2}\int_0^1\dd x\, x(1-x) \left\{ \frac{\ii \pi}{2\sqrt{-q}} + \frac{1}{\sqrt{q}} \arctan\left(\frac{p_0 x + \ii \Gamma }{\sqrt{q}}\right) -\frac{1}{\sqrt{Q}} \arctan\left(\frac{P_0 x - \ii \Gamma }{\sqrt{Q}}\right)\right\},\nonumber
\end{eqnarray}
where $q= p^2 (1-x) x -m^2$, $Q= P^2 (1-x) x - m^2$, $P^2= P_0^2 - \pB^2$, and $P_0 = p_0 + 2\ii \Gamma$. Note that the integrals over $x$ in (\ref{albeta}) are convergent.

In the limit $\Gamma\to 0$ our results agree with previous calculations,
\begin{equation}
\Pi^{ab}_{\mathrm{PV}}(\Gamma=0)=-\frac{e^2}{\pi}\left( \delta^{ab}+\frac{p^ap^b}{p^2}\right) \frac{2m|p|-(p^2+4m^2)\mathrm{arctanh}(|p|/2m)}{2|p|},\label{PiGam0}
\end{equation}
 see \cite{Appelquist:1986qw} and also \cite{Bordag:2009fz}.

\section{Conductivities}\label{sec:Dru}
From now on we restore the dependence of polarization tensor on the Fermi velocity according to Eq.\,\eqref{vF1}.

In this Section, we consider the conductivity tensor defined as
\begin{equation}
\sigma^{ab}=\frac{\Pi^{ab}_{\mathrm{PV}}}{\ii p_0}. \label{defcond}
\end{equation}
For vanishing spatial momenta, $p_a=0$, only a scalar conductivity $\sigma(p_0)$ entering the tensor conductivity as
\begin{equation}
\delta^{ab}\sigma(p_0)=\sigma^{ab}\vert_{p_a=0}\label{sigma}
\end{equation}
is important.
It is convenient to measure $\sigma(p_0)$ in the units of universal conductivity of graphene $\sigma_{\mathrm{gr}}=e^2/4$ which is nothing else than the conductivity of a pristine graphene with $\Gamma\to +0$ at $m\to 0$. In terms of the form factors from the previous Section, $\sigma(p_0)=\alpha(p_0,0)/(\ii p_0)$. For arbitrary values of the parameters $\sigma(p_0)$ can only be evaluated numerically. However, an expansion in $p_0$ for $p_0\ll \Gamma$ and $p_0\ll m$ can be done analytically,
\begin{equation}
\frac{\sigma}{\sigma_{\mathrm{gr}}}=\frac{4}{\pi^2} \left\{\frac{2 \Gamma ^2}{\Gamma ^2+m^2}  + \frac{\ii p_0}{3m} \left(\frac{\Gamma  m \left(\Gamma ^2-5 m^2\right)}{\left(\Gamma ^2+m^2\right)^2} + 2 \arctan\left(\frac{\Gamma }{m}\right) - \pi \right)+\mathcal{O}(p_0^2)\right\}.\label{sigp0}
\end{equation}
At $p_0=0$ this equation coincides with a relation obtained in \cite{Gorbar:2002iw}.

Let us check whether at small frequencies $p_0$ the conductivity can be approximated by the Drude formula
\begin{equation}\label{eq:Drude0}
	\frac{\sigma_\Dr}{\sigma_{\mathrm{gr}}} = \frac{s \gam}{\ii p_0 + \gam }=\frac{s \gam^2}{p_0^2 + \gam^2} - \ii \frac{s p_0 \gam}{p_0^2 + \gam^2},
\end{equation}
where $\gam$ has the meaning of inverse scattering rate (similarly to $\Gamma$) while $s$ defines the Drude weight. This is a phenomenological formula valid mostly for metals. There is no profound reason why it should describe the conductivity of graphene. However, this is an interesting check.

\begin{figure}
	\centering
	\includegraphics[width=0.4\linewidth]{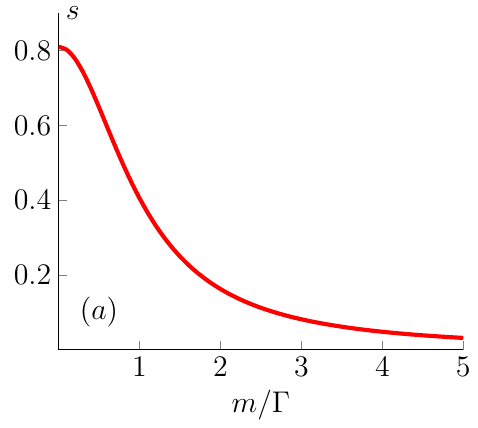}\includegraphics[width=0.4\linewidth]{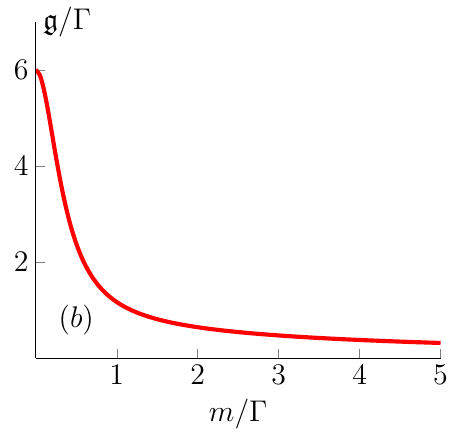}
	\caption{The plots of the parameters of Drude model given by Eq.\,\eqref{eq:Drude2}}
	\label{fig:td}
\end{figure}

\begin{figure}
	\centering
	\includegraphics[width=0.4\linewidth]{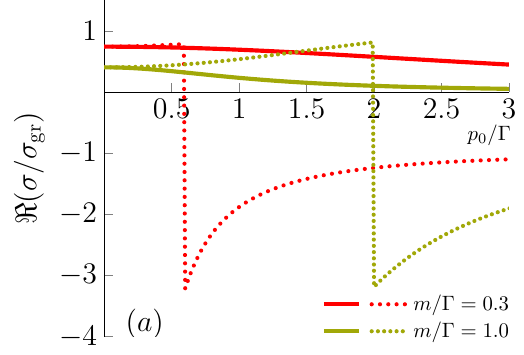}\hspace{2em}\includegraphics[width=0.4\linewidth]{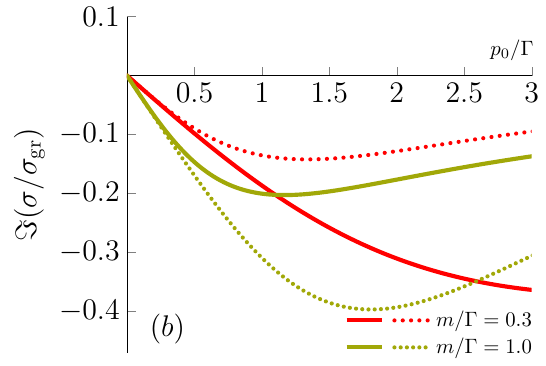}
	\caption{Real (a) and imaginary (b) parts of the conductivity as functions of $p_0/\Gamma$. The solid lines are for Drude model \eqref{eq:Drude0} with parameters \eqref{eq:Drude2}. The dashed lines are the conductivity derived from the polarization tensor with $\pB = 0$}
	\label{fig:ritd}
\end{figure}

By identifying the leading terms in the small $p_0$ expansion of (\ref{eq:Drude0}) with (\ref{sigp0}) we obtain
\begin{equation}\label{eq:Drude2}
	s = \frac{8 \Gamma ^2}{\pi^2 (\Gamma ^2+m^2)},\qquad \gam = -\frac{6 \Gamma ^2 m}{(\Gamma ^2+m^2)\left(\frac{\Gamma  m \left(\Gamma ^2-5 m^2\right)}{\left(\Gamma ^2+m^2\right)^2} + 2 \arctan\left(\frac{\Gamma }{m}\right) - \pi \right)}.
\end{equation}
This quantities are plotted at Fig.\,\ref{fig:td} while the real and imaginary parts of $\sigma$ are depicted at Fig.\,\ref{fig:ritd} in comparison with the Drude formula. 
Somewhat surprisingly, the Drude scattering rate $\gam$ may be considerably larger or considerably smaller than the scattering rate $\Gamma$ of the quasiparticles. By looking at Fig.\,\ref{fig:ritd}, we see that the agreement between imaginary parts of $\sigma$ and $ \sigma_\Dr$ is fairly good at low frequencies though the deviations become larger at higher frequencies. The real parts show qualitatively different behaviour. $\Re \sigma$ increases at low frequencies while $\Re \sigma_\Dr$ decreases. Besides, $\Re \sigma_\Dr$ does not have a jump at the threshold of pair creation $p_0=2m$ which is a characteristic feature of the conductivity of graphene. We conclude that the Drude formula fails to give a reasonable approximation to the conductivity of graphene with impurities in the range of frequencies which are relevant for the Casimir effect. The quality of approximation may be improved by using the Drude--Lorentz formulas for conductivities \cite{Drosdoff2010}.

For $\pB\neq \mathbf{0}$, it is convenient to use two different conductivities,
\begin{equation}
\sigma_\te = \frac{\alpha}{\ii p_0} \quad\mbox{and}\quad \sigma_\tm = \frac{\alpha + \beta}{\ii p_0}.\label{sitetm}
\end{equation}
As we will see in the next Section, these conductivities describe the reflection of TE and TM waves, respectively, on the surface of graphene. We will also see that TM modes give a dominant contribution to the Casimir interaction. 

\begin{figure}
	\centering
	\includegraphics[width=0.4\linewidth]{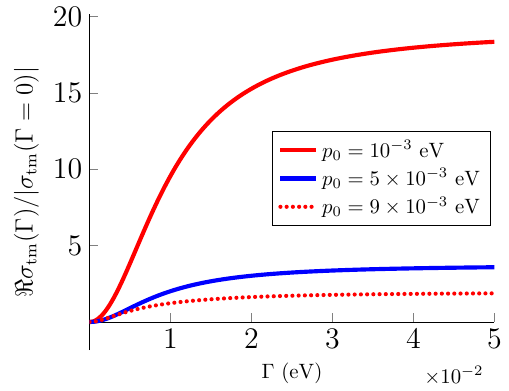}
	\caption{The normalized conductivity $\Re \sigma_\tm(\Gamma)/|\sigma_\tm(\Gamma=0)|$ as a function of $\Gamma$ for $m=0.01$ eV, $\pB=0$, and three different values of frequency $p_0$.}
	\label{fig:cond}
\end{figure}

We would like to discuss a property of the conductivities which will be important for the analysis of Casimir effect. When $\Gamma=0$ and the $\tp$ is much smaller than $m$, the imaginary parts conductivities are small while the real part vanishes, see (\ref{PiGam0}). The smallness is a direct consequence of the Pauli--Villars subtraction which ensures that the polarization tensor vanished at $m\to\infty$. Since $\Gamma$ influences conductivities, switching on $\Gamma$ at $\tp<2m$ should lead to an increase of $\Re \sigma_{\tm}$. One should only find whether this increase is large or small. Our numerical results are presented on Fig.\,\ref{fig:cond} where   $\Re \sigma_\tm(\Gamma)$ divided by the absolute values of $\sigma_\tm$ at $\Gamma=0$ is depicted as a function of $\Gamma$ for three different frequencies $p_0$ and a fixed $m=0.01$ eV. For $p_0$ much smaller than $2m$ there is a strong enhancement of the conductivity. This enhancement becomes weaker and disappears as $p_0$ approaches $2m$.

\section{The Casimir energy}\label{sec:Cas}
Let us consider two parallel infinite planes in the vacuum separated by a distance $a$. One plane will be always occupied by graphene, while the other plane will be either graphene or an ideal conductor. The Casimir energy per unit surface for two interacting plane surfaces $I$ and $II$ is given by the Lifshitz formula (see, for example, \cite{Bordag:2009zzd})
\begin{equation}
\mathcal{E}^{I,II}= \int \frac{\dd^2 \pB}{(2\pi)^3} \int_0^\infty \dd\xi \left[\ln \left(1 - e^{-2 a p_E} r^{(I)}_\te r^{(II)}_\te\right) + \ln \left(1 - e^{-2 a p_E} r^{(I)}_\tm r^{(II)}_\tm \right)\right],\label{Lifshitz}
\end{equation}
where the integrations are done over the spatial momentum $\pB$ parallel to the plates and over the imaginary frequency $\xi$, $p_0=\ii \xi$, and $p_E=\sqrt{\xi^2+\pB^2}$. In (\ref{Lifshitz}), $r_{\te,\tm}^{I,II}$ denote the reflection coefficient for TE and TM waves on the first and second surfaces, respectively. These coefficients can be expressed through the polarization tensor as \cite{Fialkovsky:2011pu}
\begin{equation}
	r_\te = -\left(1 + \frac{2p_E}{\xi \sigma_\te}\right)^{-1} , \ r_\tm =  \left(1 + \frac{2\xi}{p_E\sigma_\tm}\right)^{-1}, \label{rtetm}
\end{equation}
where $\sigma_{\te,\tm}$ have been defined above in Eq.\ (\ref{sitetm}). After continuation to the imaginary frequencies they read
$
\sigma_\te = -{\alpha}/{\xi}$, $\sigma_\tm = -{(\alpha+\beta)}/{\xi}$.	

For an ideal conductor $r_\tm =-r_\te =1$ which can be obtained from (\ref{rtetm}) by taking the limits $\sigma_{\te,\tm}\to\infty$. The Casimir energy density for two ideal conductors reads
\begin{equation}
\mathcal{E}_{\mathrm{id}} = - \frac{\pi^2}{720 a^3}. \label{Eid}
\end{equation}

We like to mention that Eq.\ (\ref{Lifshitz}) is valid only if the reflection matrix is diagonal in the TE-TM basis. Otherwise, the Lifshitz formula is a bit more complicated, see e.g., \cite{Fialkovsky:2018fpo}.

Since the reflection coefficients depend on $|\pB|$ rather than on individual components of $\pB$ we can integrate over the angular coordinates in the Lifshitz formula (\ref{Lifshitz}) which boils down to the replacement $\dd^2 \pB\to (2\pi) |\pB|\dd |\pB|$. Next, we introduce new variables, $y$ and $z$, by the formulas $\pB^2 = p_E^2 (1-y^2)$, $\xi = p_E y$, and $p_E \to z/a$. In these new variables,
\begin{equation}\label{eq:eggf}
	\mathcal{E}^{I,II} = \frac{1}{a^3}\int_0^\infty \frac{z^2 \dd z}{(2\pi)^2} \int_0^1 \dd y \left[\ln \left(1 - \frac{e^{-2 z}}{\left(1 + \frac{2}{y \sigma_\te^{I}}\right) \left(1 + \frac{2}{y \sigma_\te^{II}}\right)}\right) + \ln \left(1 - \frac{e^{-2 z}}{\left(1 + \frac{2y}{\sigma_\tm^{I}}\right) \left(1 + \frac{2y}{\sigma_\tm^{II}}\right)} \right)\right],
\end{equation}

One can notice that  apart from an overall factor of $1/a^3$ the whole dependence of Casimir energy (\ref{eq:eggf}) on the distance resides in the conductivities where it appears in the combinations 
\begin{equation}
\tilde{m} = \frac{am}{z},\qquad \tilde{\Gamma} = \frac{a\Gamma}{z}.\label{tilmGa}
\end{equation}
Explicit formulas for $\sigma_\te$ and $\sigma_\tm$ are quite long and deferred to the Appendix \ref{sec:long}, see (\ref{eq:etas2}). The limit $a\to 0$ is equivalent to taking $m\to 0$ and $\Gamma\to 0$ so that we have the case gapless pristine graphene studied in \cite{Bordag:2009fz} with $\mathcal{E}\sim 1/a^3$. 

To analyze the opposite limit of large distance we have to take $\tilde{m} \to \infty$ and $\tilde{\Gamma} \to \infty$ while keeping the fraction $\tilde{\Gamma}/\tilde{m}=\Gamma/m$ fixed. In this limit, $\sigma_\te$ and $\sigma_\tm$ behave identically,
\begin{align}
\frac{\sigma_{\te}}{\sigma_\gr},\  \frac{\sigma_{\tm}}{\sigma_\gr}\to & \frac{8 \tilde{\Gamma}^2}{\pi ^2 \left(\tilde{\Gamma} ^2+\tilde{m}^2\right)} = \frac{8 }{\pi ^2}G , \label{ssGam}
\end{align}
where
\begin{equation}
	G \equiv \frac{\Gamma^2}{m^2 + \Gamma^2} \leq 1.
\end{equation} 
Thus, we have the case of constant conductivities analysed in \cite{Khusnutdinov:2014:Cefswcc}. The Casimir energy decays as $1/a^3$ and
\begin{align}
	&\frac{\mathcal{E}^{\mathrm{g,g}}_\tm}{\mathcal{E}_{\mathrm{id}}}\approx 4.3\times 10^{-3}G,\qquad
	\frac{\mathcal{E}^{\mathrm{g,g}}_\te}{\mathcal{E}_{\mathrm{id}}}\approx 1.3\times 10^{-5}G^2,\\
	&\frac{\mathcal{E}^{\mathrm{g,id}}_\tm}{\mathcal{E}_{\mathrm{id}}}\approx 2\times 10^{-2}G,\qquad \frac{\mathcal{E}^{\mathrm{g,id}}_\te}{\mathcal{E}_{\mathrm{id}}}\approx  2.1\times 10^{-3} G, \label{asgid}
\end{align}
where the superscripts ${\mathrm{g,g}}$ and ${\mathrm{g,id}}$ mean the interaction between two identical sheets of graphene and between graphene and an ideal conductor, respectively. We normalized the results to the Casimir energy of two ideal conductors (\ref{Eid}) and separated the contributions of TE and TM modes. As usual \cite{Fialkovsky:2011pu}, the contributions of TE modes are much smaller than the TM contributions.

For $\Gamma=0$ the terms on the right hand sides of (\ref{ssGam}) vanish and the leading terms in the $a\to \infty$ expansion read
\begin{equation}
	\frac{\sigma_\te}{\sigma_\gr}  = \frac{4 z{((1-y^2) \vf^2 + y^2)}}{3\pi y (a m)} ,\qquad	\frac{\sigma_\tm}{\sigma_\gr} = \frac{4 z{y}}{3\pi (a m)}.
\end{equation} 
As a result, the Casimir energy of two graphene sheets behave at large distances as $\mathcal{E} \sim 1/(a^3 (am)^2) $ while for the system of a  perfect metal and graphene the asymptotic behaviour is $\mathcal{E} \sim 1/(a^3 (am))$, in agreement with \cite{Bordag:2009fz}.

The rest of the analysis will be done numerically.
For orientation, we present the value of $\Gamma=4.4\times 10^{-3}$ eV which gives the best fit \cite{Fialkovsky:2012ee} between the Quantum Field Theory polarization tensor and the Faraday rotation experiment \cite{Crassee_2010} at $7$ T.\footnote{For a magnetic field of $3$T  different values of the parameters were obtained as well as a much worse fit. This can be attributed to some specific physics in particular samples of graphene used in the experiment, see \cite{Crassee_2012}.} The separation $a$ used in the Casimir experiments with graphene \cite{Banishev:2013,Liu:2021ice} is between $200$ nm and $2000$ nm. The value for $m$ for the graphene sample used in the Casimir experiment \cite{Liu:2021efo} was $\sim 0.15$ eV. The presence of a damping factor $e^{-2ap_E}$ in the Lifshitz formula (\ref{Lifshitz}) indicates that the region of relevant Euclidean momenta is restricted by the inequality $p_E\lesssim (2a)^{-1}$. E.g., for $a=500$ nm one has $(2a)^{-1}\simeq 0.2$ eV. 

\begin{figure}
	\centering
	\includegraphics[width=0.3\linewidth]{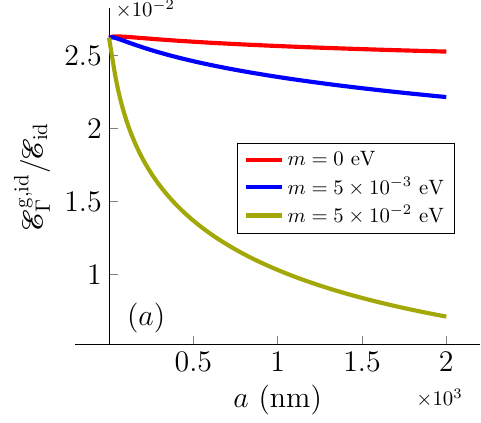}\includegraphics[width=0.3\linewidth]{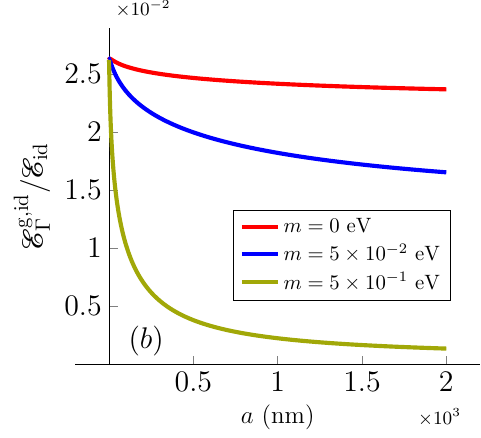}\includegraphics[width=0.3\linewidth]{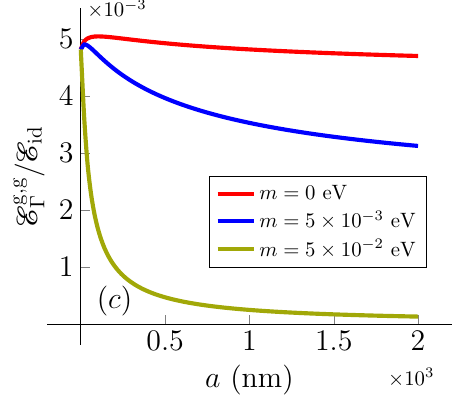}
	\caption{The Casimir energy density as a function of the distance normalised $\mathcal{E}_{\mathrm{id} }$ (Eq. (\ref{Eid})) for the system of graphene and an ideal conductor (on (a) and (b)) and for two identical graphene sheets (c). The impurity scattering rate is  $\Gamma=5\times 10^{-3}$ eV on (a) and (c), and $\Gamma=5\times 10^{-2}$ eV on (b).}
	\label{fig:Cas}
\end{figure}

\begin{figure}
	\centering
	\includegraphics[width=0.4\linewidth]{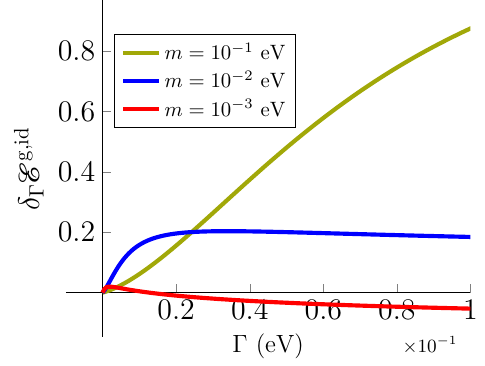}
	\caption{The plots of relative variation of the Casimir energy density (\ref{relvarE}) at $a=500$nm as a function of $\Gamma$ for three values of the mass.}
	\label{fig:Casmore}
\end{figure}

The Casimir energy density normalized to the Casimir energy density between two ideal conductors is plotted on Fig.\,\ref{fig:Cas}. By comparing Fig.\,\ref{fig:Cas}(a,b) with Fig.\,\ref{fig:Cas}(c) we see that the Casimir interaction of graphene with graphene behaves qualitatively similarly to the interaction of graphene with an ideal conductor though the former is about 5 times weaker than the latter. Thus, in what follows we will only consider the interaction of graphene with an ideal conductor. On Figs.\,\ref{fig:Cas} (a) and (b) we see that the Casimir interaction is damped by the mass exactly as it happens for $\Gamma=0$, see \cite{Bordag:2009fz}. The line for $m=0$ appears slightly upper for $\Gamma=0.005$ eV than for $\Gamma=0.05$ eV, while the line for $m=0.05$ eV passes significantly lower on Fig.\,\ref{fig:Cas}(a) than on Fig.\,\ref{fig:Cas}(b). Thus, the effect of $\Gamma$ strongly depends on the mass. To see this effect more clearly, we depicted on Fig.\,\ref{fig:Casmore} the relative variation of Casimir energy density
\begin{equation}
\delta_\Gamma \mathcal{E}^{\mathrm{g},\mathrm{id}}=\frac{\mathcal{E}^{\mathrm{g},\mathrm{id}}(\Gamma)-\mathcal{E}^{\mathrm{g},\mathrm{id}}(\Gamma=0)}{\mathcal{E}^{\mathrm{g},\mathrm{id}}(\Gamma=0)}\label{relvarE}
\end{equation}
at a fixed separation $a=500$ nm and three different values of the mass. 

 The large relative enhancement of the Casimir effect for $m=0.1$ eV can be explained by the observation made at the end of the previous section that turning on $\Gamma$ leads to a significant enhancement of $\Re \sigma_\tm$ for the frequencies $p_0\ll m$. For a larger mass the interval of frequencies where the enhancement of conductivity takes place becomes larger which is translated to the enhancement of Casimir interaction. Since the Casimir effect is an integral effect containing competing contributions from all frequencies and tangential momenta it is impossible to predict the amplitude of the enhancement basing on this type of arguments. For the same reason, it is hard to give a precise explanation to the very moderate damping of Casimir force by $\Gamma$ at very small  masses like $m=10^{-3}$ eV at Fig.\ \ref{fig:Casmore}. We can only suppose that at the absence of the mechanism described above the impurity scattering should reduce the conductivities and thus the Casimir interaction according to the high-school physics intuition. For an intermediate mass, $m=10^{-2}$ eV, we see an intermediate behaviour.

Another characteristic feature of the Casimir interaction in the presence of $\Gamma$ is a $1/a^3$ behaviour of the Casimir energy at large distances. However, the asymptotic values (\ref{asgid}) are reached at very large $a$. For example, for $m=\Gamma=0.01$ eV at $a=10^6$ nm the difference between exact and asymptotic values is still at the 10\% level. For some other choices of $m$ and $\Gamma$ the agreement may be reached at somewhat smaller distances which are anyway beyond the distances used at experiments.

%\section{Non-dispersive approximation}\label{sec:nondis}

\section{Conclusions}\label{sec:concl}
In this paper, we studied the Casimir interaction of graphene containing impurities described by a scattering rate $\Gamma$. We evaluated the polarization tensor of quasiparticles in the Pauli--Villars subtraction scheme and computed the corresponding conductivities. Next, we used the conductivities in the Lifshitz formula to calculate the Casimir energy density for various values of parameters. Our main message is that the presence of impurities can considerably enhance the Casimir interaction for large values of the mass (about $m=0.1$ eV) and somewhat damp the Casimir interaction for near zero mass. Thus, the non-zero impurity scattering rate $\Gamma$ should be taken into account in the analysis of precision Casimir experiments with graphene. To do this however one should also take into consideration the chemical potential and temperature since both influence in an essential way the Casimir force \cite{Fialkovsky:2011pu,Bordag:2015zda,Liu:2021efo}. This will be our next task which we are going to take up in the near future.

An additional motivation for considering the influence of impurities jointly with the temperature comes from the old discussion on the Nernst theorem in the Casimir physics \cite{Bezerra:2004zz,Brevik:2007wj,Bezerra:2005hc}, see \cite{Mostepanenko:2021cbf} for a recent review. The violation of Nernst theorem was blamed on the relaxation rate parameter present in the Drude formula. It is interesting to check what happens if this rate appears in a different model.

In \cite{Holanda:2023joy} an unconventional version of the Pauli--Villars subtraction scheme in the presence of impurities was suggested. It would be interesting to study the influence of this scheme on Casimir interaction.

Finally, we would like to mention an approach to the Casimir interaction of graphene based on the QFT effective action \cite{Bordag:2009fz,Farias:2016uls}. This approach is equivalent to a fine structure constant expansion of the Lifshitz formula.

\begin{acknowledgments} 
We are grateful to Ignat Fialkovsky for previous collaboration and discussions on the impurities in graphene. This work was supported in parts by the S\~ao Paulo Research Foundation (FAPESP) through the grants 2021/10128-0 (D.V.) and 2022/08771-5 (N.K.), and by the National Council for Scientific and Technological Development (CNPq), grant 304758/2022-1 (D.V). 
\end{acknowledgments}

\appendix
\section{Long but useful formulas}\label{sec:long}
Here we present explicit form of the conductivities after the changes of variables made above Eq.\ (\ref{eq:eggf}) and in (\ref{tilmGa}).
\begin{align}
	\frac{\sigma_\te}{\sigma_\gr}  = \frac{8}{\pi^2 y}\int_0^1& \dd x \left\{\frac{ \pi{((1-y^2) \vf^2 + y^2)} (1-x) x}{\sqrt{\tilde{m}^2 + (1-x) x((1-y^2) \vf^2 + y^2)}}\right.\ann
	& - \frac{2{((1-y^2) \vf^2 + y^2)} (1-x) x }{\sqrt{\tilde{m}^2 + (1-x) x((1-y^2) \vf^2 + y^2)}} \arctan\left(\frac{\tilde{\Gamma} + y x }{\sqrt{\tilde{m}^2 + (1-x) x((1-y^2) \vf^2 + y^2)}}\right)\ann
	&-\left.  \frac{2{((1-y^2) \vf^2 +(y + 2\tilde{\Gamma})^2)} (1-x) x }{\sqrt{\tilde{m}^2 + (1-x)x ((1-y^2) \vf^2 +(y + 2\tilde{\Gamma})^2)}} \arctan\left(\frac{\tilde{\Gamma} - x(y + 2\tilde{\Gamma}) }{\sqrt{\tilde{m}^2 + (1-x)x ((1-y^2) \vf^2 +(y + 2\tilde{\Gamma})^2)}}\right)\right.\ann
	&+ \left. \frac{1}{2} y (1-2 x) \ln \left((y x + \tilde{\Gamma} )^2 + \tilde{m}^2 + (1-x) x((1-y^2) \vf^2 + y^2)\right)\right.\ann
	&-\left. \frac{1}{2} (y + 2 \tilde{\Gamma}) (1-2 x) \ln \left( (\tilde{\Gamma} - (y + 2\tilde{\Gamma}) x)^2 + \tilde{m}^2 + (1-x)x ((1-y^2) \vf^2 +(y + 2\tilde{\Gamma})^2)\right)\right\},\ann
	\frac{\sigma_\tm}{\sigma_\gr} = \frac{8}{\pi^2 y}\int_0^1& \dd x \left\{\frac{ \pi{y^2} (1-x) x}{\sqrt{\tilde{m}^2 + (1-x) x((1-y^2) \vf^2 + y^2)}} \right.\ann
	&- \frac{2{y^2} (1-x) x }{\sqrt{\tilde{m}^2 + (1-x) x((1-y^2) \vf^2 + y^2)}} \arctan\left(\frac{\tilde{\Gamma} + y x }{\sqrt{\tilde{m}^2 + (1-x) x((1-y^2) \vf^2 + y^2)}}\right)\ann
	&-\left.  \frac{2{(y + 2\tilde{\Gamma})^2} (1-x) x }{\sqrt{\tilde{m}^2 + (1-x)x ((1-y^2) \vf^2 +(y + 2\tilde{\Gamma})^2)}} \arctan\left(\frac{\tilde{\Gamma} - x(y + 2\tilde{\Gamma}) }{\sqrt{\tilde{m}^2 + (1-x)x ((1-y^2) \vf^2 +(y + 2\tilde{\Gamma})^2)}}\right)\right.\ann
	&+ \left. \frac{1}{2} y (1-2 x) \ln \left((y x + \tilde{\Gamma} )^2 + \tilde{m}^2 + (1-x) x((1-y^2) \vf^2 + y^2)\right)\right.\ann
	&-\left. \frac{1}{2} (y + 2 \tilde{\Gamma}) (1-2 x) \ln \left( (\tilde{\Gamma} - (y + 2\tilde{\Gamma}) x)^2 + \tilde{m}^2 + (1-x)x ((1-y^2) \vf^2 +(y + 2\tilde{\Gamma})^2)\right)\right\}. \label{eq:etas2}
\end{align}

%\bibliography{graphene}

\end{document}